# Comparison of D-Wave Quantum Annealing and Classical Simulated Annealing for Local Minima Determination

Yaroslav Koshka, M.A. Novotny

*Abstract*—**In this work, Restricted Boltzmann Machines (RBMs) trained with different numbers of iterations were used to provide a large diverse set of energy functions each containing many local valleys (LVs) with different energies, widths, escape barrier heights, etc. They were used to verify the previously reported possibility of using the D-Wave quantum annealer (QA) to find potentially important LVs in the energy functions of Ising spin glasses, Markov Random Fields and related problems that may be missed by classical searches. Instead of the technique of the previous work, similar to the one used in training graphical models with Contrastive Divergence (i.e., initiating the Markov chain from each of the training patterns in the dataset), which is useful for time-efficient sampling in machine learning, this work utilized extensive simulated annealing (SA) with an increasing duration in an attempt to find classically as many LVs as possible regardless of the computational cost. SA was conducted long enough to ensure that the number of SA-found LVs approaches that and eventually significantly exceeds the number of the LVs found by a single call submitted to the D-Wave. Even after a prohibitively long SA search, as many as 30-50% of the D-Wave-found LVs remained not found by the SA. In order to establish if those LVs that are found only by the D-Wave represent potentially important regions of the configuration space, they were compared to those that were found by both techniques with respect to different properties of the corresponding LVs. While the LVs found by the D-Wave but missed by SA predominantly had higher energies and lower escape barriers, there was a significant fraction having intermediate values of the energy and barrier height, even after the longest classical search attempted in this work. With respect to most other important LV parameters, the LVs found only by the D-Wave were distributed in a wide range of the parameters' values. Finally, in an attempt to explain which LVs could not be found easily by the SA, it was established that for large or small, shallow or deep, wide or narrow LVs, the LVs found only by the D-Wave are distinguished by a few-times smaller size of the LV basin of attraction (BoA), which was estimated as the number of higher-energy states sampled above the height of the smallest escape barrier. Apparently, the size of the BoA is not or at least is less important for QA search compared to the classical search, allowing QA to easily find many potentially important (e.g., wide and deep) LVs missed by even prohibitively lengthy classical searches.**

## I. Introduction

Thanks to the introduction of the first commercial quantum annealing computer by the D-Wave Systems, Inc [1], adiabatic quantum annealers (QAs) are aggressively investigated for a vast variety of discrete optimization tasks. For example, the D-Wave was used to classify Higgs-boson-decay signals [2], graph partitioning using D-Wave was investigated and was shown to be comparable to other state-of-the-art methods [3], the D-Wave was used for classification of DNA sequences according to their binding affinities [4], for matrix factorization [5], and for many other applications.

There is a strong interest in adiabatic QA for machine learning (ML). For example, the quantum hardware was used for training deep neural networks (DNNs) [6]-[10]. Besides the traditional ML tasks facilitated by QA, there are promising results in using neural networks for solving the quantum many-body problems [11]. The use of a QA could further enhance the potential of the neural-network-based approach to solving quantum many-body problems.

In our previous work [12] a comparison between the D-Wave and a classical Markov Chain Monte Carlo (MCMC) was conducted to evaluate how well the samples obtained by both techniques represent the probability distribution of a Markov Random Field, and Restricted Boltzmann Machines (RBMs) in particular. Specifically, the QA and the classical samples were compared in terms of the local valleys (LVs) to which the states sampled by the two techniques belonged, which means, how many and what LVs are represented or not represented by the particular sample, and how important potentially the found and the missed LVs are for ML. For this purpose, the classical determination of the LVs was done by initiating Gibbs chains from each of the training patterns (similar to how it is done in ML during the RBM training by contrastive divergence), followed by relaxation to the bottom of the corresponding LVs to find the local minimum (LM) and thereby determine what

This material is based on research sponsored by the Air Force Research Laboratory under agreement number FA8750-18-1-0096. The views and conclusions contained herein are those of the authors and should not be interpreted as necessarily representing the official policies or endorsement, either expressed or implied, of the Air Force Research Laboratory (AFRL) or the U.S. Government. Koshka is with the *Department of Electrical and Computer Engineering, HPC2 Center for Computational Sciences, Mississippi State University, Mississippi State, MS 39762, U.S.A.* (e-mail: ykoshka@ece.msstate.edu).

M.A. Novotny is with the *Department of Physics and Astronomy, HPC2 Center for Computational Sciences, Mississippi State University, Mississippi State, MS 39762, U.S.A.* (e-mail: man40@msstate.edu).

LV this state belongs to. This kind of search is good for finding those LVs that are known to be present and the most important in trained neural networks, i.e., the LVs that form in the vicinity of the training patterns to increase the likelihood for the training data.

In Ref.[12], the classical search missed many LVs found by the D-Wave, many of which could be expected to be potentially important for ML applications. However, it could not be ruled out that many LVs missed by the classical technique would be found by a more rigorous classical search, which would eliminate or reduce the comparative advantages of the D-Wave for this task.

Proving that there is a significant number of the LVs that can be found only by the D-Wave even when using a much more rigorous than in Ref.[12] classical search would be very interesting for many applications concerned with LVs. Of course, many of those LVs can be expected to be potentially of modest-to-low importance, depending on the application (i.e., lower probability of the corresponding states due to a not low enough energy of the LM, lower stability of the corresponding states (e.g., metastable states) due to a low escape barrier from the LV, smaller number/variety of different high-probability states in the LV because of a small width of the LV and low density of states (DOS) at the bottom, etc. However, if, in fact, it turns out that the D-Wave can find a meaningful number of low-energy, large, deep and/or wide LVs missed by very rigorous classical sampling or classical LV search, this could create a unique niche for the QA in many applications in ML, physics, engineering, materials science and other disciplines dealing with a search for local extrema.

In this work, an RBM was trained with different numbers of training epochs to produce a variety of energy landscapes with very different numbers of LVs and high variety of LV energy, depth, width and DOS. Here, the term "epoch" is used to define one full cycle of training iterations, with each training pattern participating in the given training cycle only once. Multiple prolonged cycles of simulated annealing (SA) at different temperature regimes, totaling weeks of continuous processing, were conducted to find as many LVs as possible. The number and the properties of the classically found LVs were statistically compared with those found by the D-Wave on the embeddings of the same RBMs into the Chimera lattice following the approach of Refs.[13][14]. An explanation is suggested and justified for why many potentially important LVs found by the single D-Wave call could not be found by the classical search (even after a prohibitively long classical search, which finds a much higher total number of the LVs than the D-Wave).

## II. Simulated and Quantum Annealing

### A. *Classical searches for the global and local minima*

Searching across a wide range of states is important in many fields, for example in computational chemistry (e.g., search of a chemical composition), materials discovery, etc. The list of the most popular local search algorithms includes threshold accepting methods, Noising method, genetic algorithms and the classical SA. Threshold accepting algorithm [15] is a method that relies upon a non-increasing deterministic step function during the search. Noising method [16] is a simple descent algorithm in which the solution space is perturbed by adding random noise to the problem's objective function. Tabu search [17] utilizes a memory and a dynamic list of forbidden moves. Genetic algorithms [18] utilize a set of genetically inspired stochastic transition operators to transform candidate solutions into a descendent population.

SA is a local search algorithm [19] that allows escaping from LMs by utilizing hill-climbing. It is used most often for discrete, and less often for continuous, optimization problems. The algorithm employs a temperature parameter, which is gradually decreased to zero. If the algorithm is convergent, the temperature parameter reduction leads to the probability of the obtained states being concentrated on the set of globally optimal solutions. Otherwise the algorithm converges to a local optimum.

While ensuring that SA can converge to the global extrema is the usual goal of optimizing in the SA algorithms, the present work concerned with the natural property of the SA to end up in a variety of the LMs.

### B. *Adiabatic Quantum Annealing*

Adiabatic QA is used to find the global minimum of the energy function of a spin glass by employing not only thermal jumps from one LM to another over the LM separating barriers (as in the classical SA algorithm), but also quantum mechanical tunneling through the barriers [20]. The D-Wave machine was the first commercial QA [1]. It implements an Ising spin-glass model [21], which makes it suitable for a wide range of Markov Random Fields and related problems. The optimization variables $s_j$ are represented by the qubits. Depending on the hardware architecture, couplings are provided between specific qubits. In addition, each qubit has an associated bias fields $h_j$. The weights of the couplings $J_{ij}$ and the bias fields $h_j$ are assigned values to represent a particular spin glass problem described by Eq. (1).

$$E(s) = -\sum_{i=1}^{N-1}\sum_{j=i+1}^{N} J_{ij} s_i s_j - \sum_{j=1}^{N} h_j s_j \,. \quad (1)$$

The ground state (GS) of the Ising spin glass is described by the states of all $s_i$ that correspond to the minimum of the energy function of Eq. (1) [22]. The physics of the GS determination by the D-Wave QA has been extensively investigated [23]-[25].

The Hamiltonian for an adiabatic QA can be represented as:

$$\mathcal{H}_{AQC}(s) = A(s)\mathcal{H}_{\text{Driver}} + B(s)\mathcal{H}_{\text{Problem}} \quad (2)$$

where $\mathcal{H}_{\text{Driver}}$ and $\mathcal{H}_{\text{problem}}$ are the driver and the problem Hamiltonians respectively, $s$ is the anneal fraction and $A(s)$ and $B(s)$ are the anneal functions. Initially, at $s = 0$, $A(s) >> B(s)$, and there is only $\mathcal{H}_{\text{Driver}}$ term. The $n$ qubits are placed into the quantum state that is an equal superposition of all $2^n$ possible states. Following that, $A(s)$ and $B(s)$ are changed adiabatically from $s = 0$ to $s = 1$. At the end (when $s = 1$), $A(s) << B(s)$, and the Hamiltonian of the system becomes the problem Hamiltonian $\mathcal{H}_{\text{Problem}}$. A measurement on the quantum superposition state of system at $s = 1$ should provide one of the GSs of the $\mathcal{H}_{\text{Problem}}$, if the quantum adiabatic theorem holds during the anneal.

In reality, thanks to the unavoidable deviations from the adiabatic evolution, different solutions may come from multiple repetitions of the QA run, including not only the GS but also a multitude of the excited states, possibly belonging to a multitude of different LVs.

Up to 10,000 repetitions of the quantum annealing done almost instantaneously (1000 results within tens of milliseconds) provide us with what could be considered a big "sample" from the model probability distribution. Naturally, this sample does not follow the Boltzmann distribution, or at least not the Boltzmann distribution at a temperature desirable for many applications in physics and ML. However, investigations of the ways to convert the distribution obtained from the D-Wave into the desirable form of the distribution have been reported [8]. In our previous work [12]-[13], the motivation was to investigate if a "sample" from the D-Wave is likely to reflect most of the important LVs that should be present in a sample that follows the Boltzmann distribution. In the present work, however, the main focus is specifically on those states (and LVs) that have been shown in our previous work to be consistently missed by the classical search but found by the QA.

## III. Methods

### A. RBM training and embedding within the D-Wave lattice

In this work, the D-Wave was not used for RBM training. Instead, the training of an RBM having 64 visible and 64 hidden units was conducted on a classical computer using the standard contrastive divergence (CD) [26]. After that, the trained RBM was embedded within the D-Wave lattice, and QA was used to obtain samples from the model distribution. The details of the CD training and embedding of the problem Hamiltonian in the quantum hardware can be found in Ref. [12][13]. The embedded spin glass problem corresponds to the $\mathcal{H}_{\text{Problem}}$ Hamiltonian in Eq.(2) and, also, is described by the Eq.(1) for the spin glass energy when N qubits are used.

### B. Investigation of the LVs in the energy function

In Ref.[12], we introduced a conservative approach to assess if adiabatic QA has a potential to offer important advantages to the task of finding LVs in the energy function of an Ising spin glass, Markov Random Fields and related problems. In particular, the RBM model probability distribution was selected as a versatile example for investigation in the previous and this work. The approach is based on comparing the D-Wave and the classical Monte Carlo (MC) samples, looking specifically at what LVs the states in each of the samples belong to. This approach allows investigating the relative importance of the LVs revealed by the D-Wave and by the classical samples by comparing properties of the found LVs. Those properties (parameters) include the RBM energies of the corresponding LMs, the depth of the most critical (for most applications) lower part of the LVs (i.e., the height of the escape barrier) $\Delta E_{act}$, a relative size of the LVs, which is expressed as the number of states $N_{LV}$ belonging to this LV that can be sampled during multiple cycles of a simulated warming (SW), a relative size $N_{low}$ of the main (the lower) part of the LVs below the lowest escape barrier, and a width-related parameter $W$ of the main lower part of the LVs. Parameters for the lower part of the LV is of particular interest because this is the region of the configuration space where the system spends most of the time at temperatures relevant for many applications and where higher-probability states of probabilistic graphical models are located.

Fig. 1 shows a schematic diagram of a cross-section of a LV, which is used to illustrate the parameters of LVs used in this work. For ML, the existence of LVs with associated basins of attraction (BoA) for each valley (e.g., as illustrated in Fig. 1) is a useful but rough model for what occurs in the high dimensional parameter space explored by MC methods. Such a mathematical model is also correct, with some reasonable approximations, for zero temperature MC where the algorithm runs downhill only until it is caught in a local minimum. At reasonably low temperatures, the model is still useful, and one can then consider the paths from one LV to another and the merging of BoAs as the temperature gets larger. Novotny and collaborators have performed work on the algorithms associated with such transitions [27]-[29]. Computational methods for classical MC systems to understand the high-dimensional transitions between LVs include Transition Path Theory (TST) [30], the string or elastic band method for calculating the most probable paths at low temperatures between LVs [31][32] for continuum systems and MC with Absorbing Markov Chains (MCAMC) [27] for discrete models.

The methods used in this work for evaluating parameters of Fig.1 for a large number of LV and with reasonable computational costs are described in Section C. It should be noted that in this work, we distinguish the entire LV with its BoA, which could be characterized, e.g., by $N_{LV}$ and by the highest energy barrier for escaping from the LV $\Delta E_{max}$, from the main (the lower) part of the LV below $\Delta E_{act}$, which includes states $N_{low}$. Normally, when speaking about the stability of the metastable states inside the given LV or about the number of the highest-probability states in the LV, the important parameters are $\Delta E_{act}$, $W$ and their DOS characterizing this lower most important portion of the LV (see Fig.1). On the other hand, considering the BoA may be important when analyzing how a local search algorithm finds or misses particular LVs.

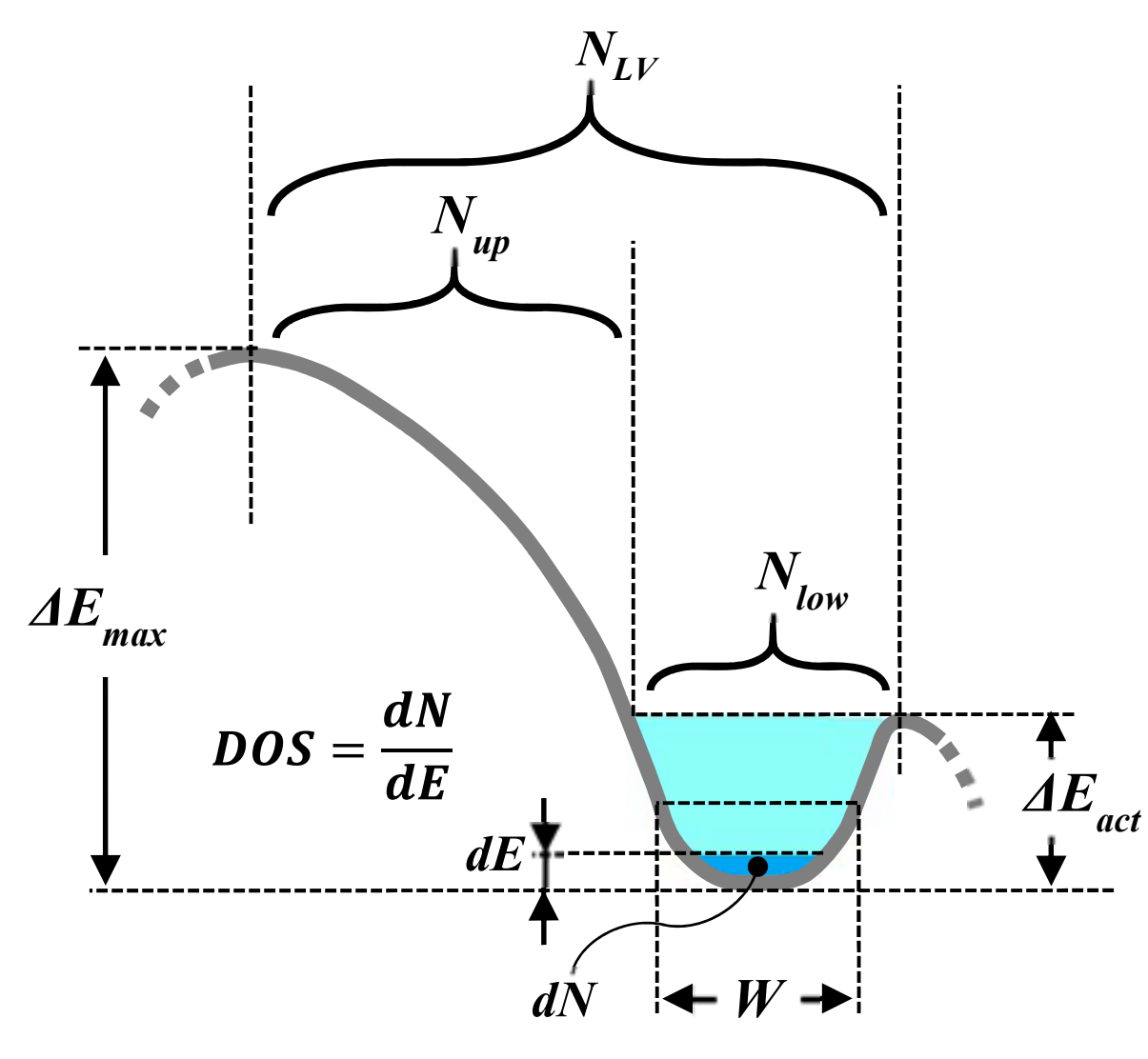

Fig.1 Schematic diagram of the profile of a LV of the energy function, illustrating parameters used in this work. The labeled shaded regions and/or curly brackets highlight different groups of states in the configuration space. $N_{LV}$ is the number of sampled states inside the LV. $N_{up}$ and $N_{low}$ are the numbers of states above and below $\Delta E_{act}$ respectively sampled from each LV. DOS is the density of states near the bottom of the LV. $\Delta E_{act}$ is the barrier height, which is the lowest barrier for escaping from the LV. $\Delta E_{max}$ the highest energy for escaping from the LV. $W$ is the width of the LV (of the major part of the LV below $E_{act}$).

The classical approach to finding the LVs of the RBM model distribution utilized SA. It was performed on the original RBM graph, i.e., the graph that was free from the representation of the RBM units by combining qubits in the D-Wave embedding. Markov chains were initiated at very high temperature to ensure nearly random initial states when the annealing starts. A series of multiple cycles having various temperature regimes, from a very fast to a very slow cooling, were conducted, with $T$ gradually reduced to zero. The selected series of those multiple cycles having different temperature regimes was then continuously repeated over a duration of multiple weeks of computer time, until the time to find any new LVs became too large, making continuation of this search procedure impractical.

In every single D-Wave call, 10,000 quantum annealing sequences were conducted in search of the lowest-energy solutions for the Hamiltonian corresponding to the embedding of the trained RBM (described in Section III.A). As a result, 10,000 solution attempts were obtained, some of which represented the same solution. This is why the number of distinct sampled states from the given D-Wave call typically was less than 10,000.

Next, the D-Wave sample was analyzed to establish to what LVs the states found by the 10,000 D-Wave anneals belonged. The vector of the visible RBM units for each distinct D-Wave solution was used as an initial state, and a SA was conducted at $T=0$ for each of those states to evolve the system downhill until the bottom of the given LV is reached.

### C. Simulated Warming and estimation of the LV parameters

Simulated warming (SW) was used to investigate the properties of the large number of LVs found by the SA and the QA. MCMC was conducted with the temperature gradually increasing from its initial value of $T=0$. The samples were collected and evaluated after each MCMC jump.

To identify those sampled states after SW that belong to the same LV from which the SW was initiated, a downhill evolution from each of the sampled states was conducted, until the corresponding LM was reached. This was done to establish if the sampled state indeed corresponds to the initial LV of the state found by the D-Wave, or if the system escaped into another LV. The number of the sampled states that remain in the same LV after the SW was used as a comparative estimate of the size of the LVs and, indirectly, the DOS.

The escape frequency for a given LV was estimated by counting and taking an inverse of the average number of the SA jumps before a state outside of the initial LV is obtained. SW was conducted at different temperatures to determine the activation energy of the escape frequency $\Delta E_{act}$, which served as a measure of the energy barrier related to the thermal stability of the high-probability states at the bottom of the LV.

A relative size of the LVs (including the BoA) was evaluated as the number of states $N_{LV}$ belonging to this LV that can be sampled at the same conditions during multiple cycles of SW, while increasing $T$ until all Markov chains start to "escape" a given LV without finding any new states belonging to the LV. Of course, far from all (or even a significant fraction of the total number of) the states belonging to a LV could be sampled. However, for the purpose of comparing different LVs, it is reasonable to assume that the average number of the sampled states is proportional to the actual total number of the states and therefore serves as a sufficiently reliable qualitative comparative estimate.

$N_{low}$ is a relative size of the most critical lower part of the LVs, where the system spends most of the time at temperatures relevant for most applications and where higher-probability states of the probabilistic graphical models are located. In this work, $N_{low}$ was somewhat arbitrary represented by those sampled states from the total $N_{LV}$ that had energy below $\Delta E_{act}$.

The width-related parameter $W$ was estimated using an approximation of a LV with a square well, and $W$ was calculated by dividing $N_{low}$ of this LV by the depth of the well $\Delta E_{act}$. Another approach relied upon determining the intercept of the Arrhenius plot with the vertical axis, which provided qualitatively similar results for the $W$ estimate.

An estimate of the density of states DOS at the bottom of LVs was used to compare the LVs with respect to the number of the available highest-probability states. The DOS was calculated by dividing the number of the bottom states $dN$ with energies in a narrow range $dE$ by the value of $dE$.

The highest temperatures used in this work were significantly higher than those used by us in Ref.[12]. At these highest temperatures, most of the Markov chains escape the LV before a state gets sampled, resulting in very infrequent finding of a new state belonging to the LV. This allowed us to probe the most remote regions of the BoA of the LV up to the highest energy for escaping from the LV $\Delta E_{max}$, and find the number of the BoA states (upper states) $N_{up}$, defined as the sampled states outside of the lower (the main) portion of the LV, which (somewhat arbitrary) were selected as having the energy higher than $\Delta E_{act}$. The values of $N_{up}$ were compared for LVs found only by the D-Wave and LVs found by both techniques, with the goal of determining the relative importance of the effect of the LV's BoA on finding or missing a given LV by each of the techniques.

## IV. Results And Discussion

Classification of the handwritten digit patterns by the RBM, reconstruction of partial patterns and pattern generation approach reported in Ref.[12] was also employed in this work to verify that the RBM embedding into the D-Wave can be used to correctly determine the lowest energy state of a complex probability distribution under a variety of constrains (e.g., clamped visible units, clamped labels, etc). Having obtained this reassurance, we then searched for LVs and compared LVs found by the QA and SA for different energy functions corresponding to different numbers of RBM training iterations.

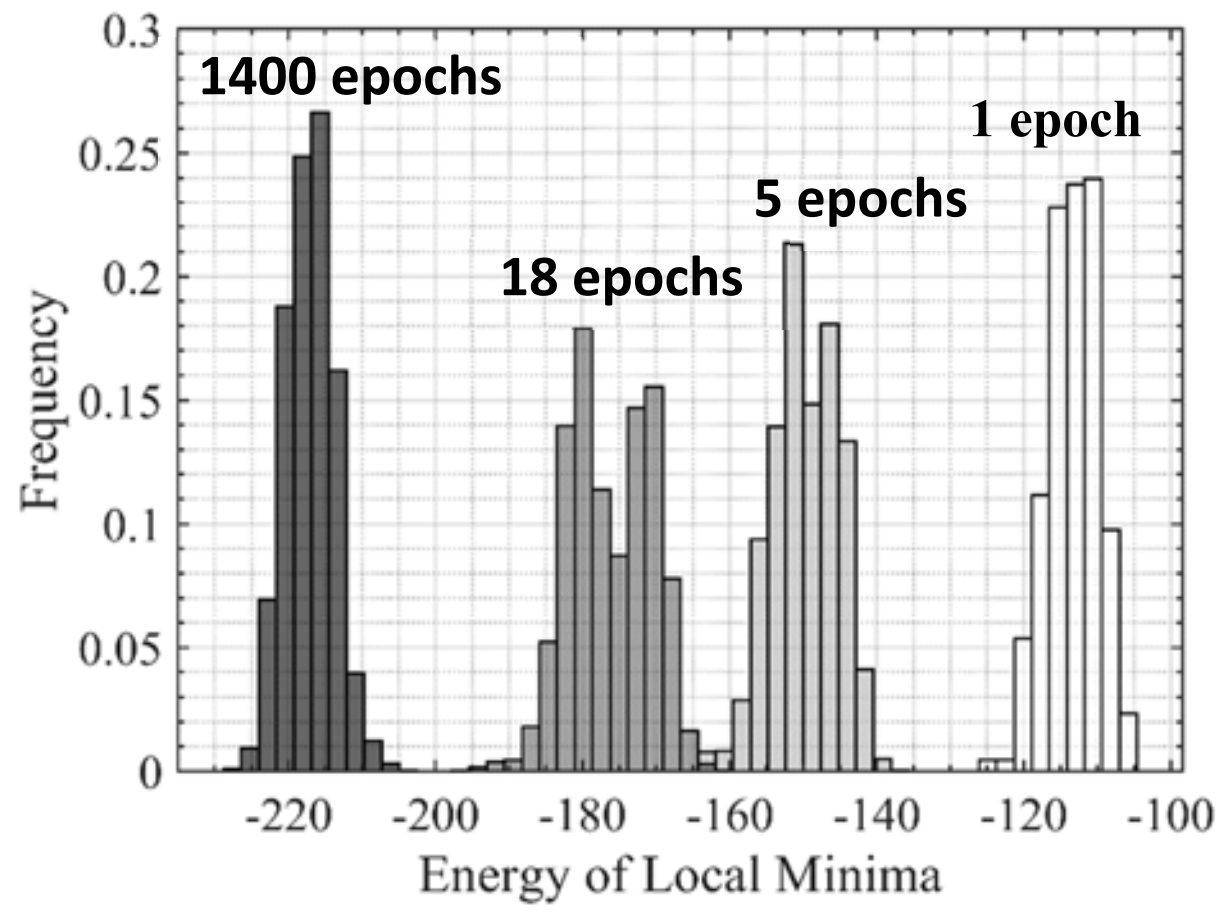

Fig.2 Histograms of the energies of the LMs found in four energy functions used in this work: RBMs trained with 1, 5, 18 and 1400 epochs. The LMs in each case were found after the 5th (the last) cycle of many SA runs starting from random initial states.

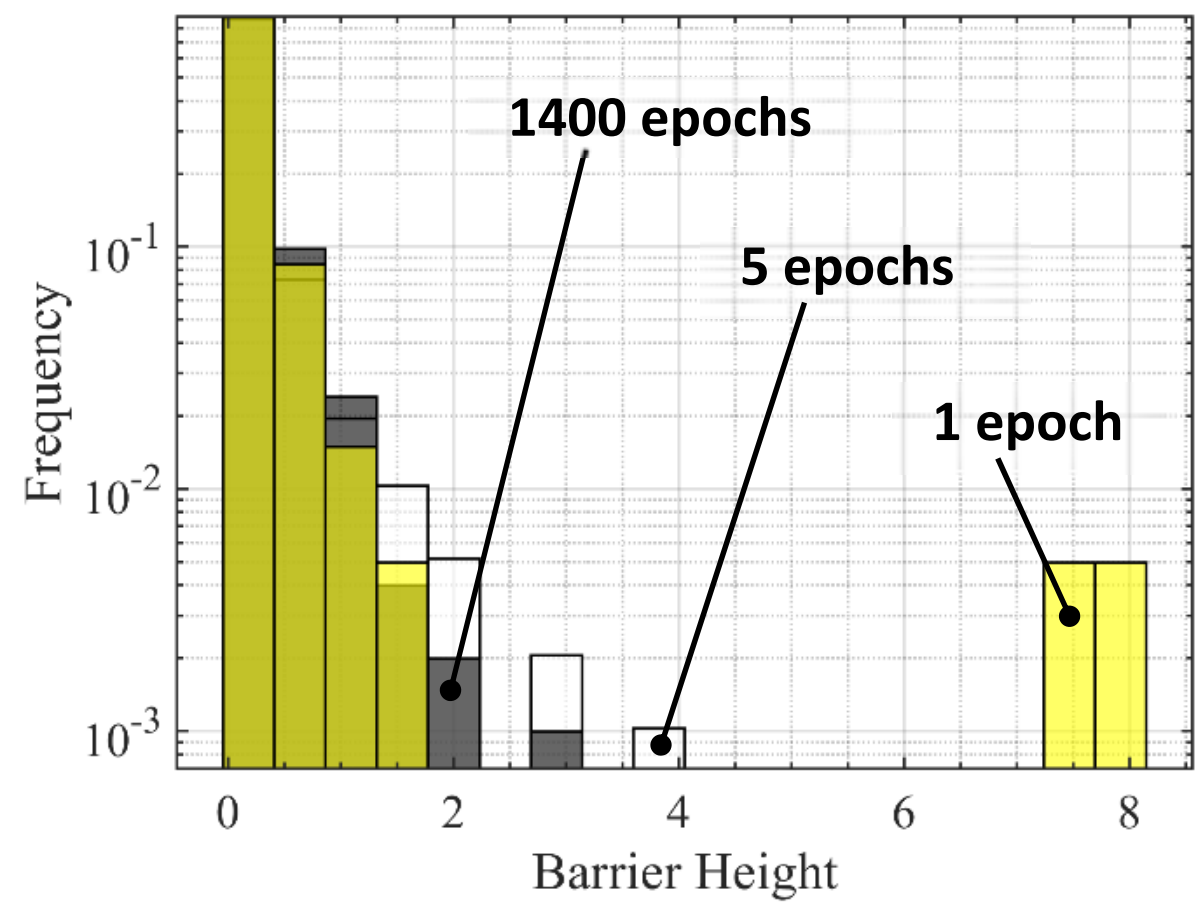

Fig.3 Histograms of the barrier heights (the escape barriers $E_{act}$) of the LVs for three (out of four shown in Fig.2) energy functions: RBMs trained with 1, 5 and 1400 epochs. The LMs here were found by relaxation at $T$=0 from each solution found by the D-Wave.

### A. The variety of the investigated energy functions

First, we look at how different the energy functions produced by different training iterations are from each other with respect to a few main parameters of the LVs.

Fig. 2 shows histograms of the RBM energies of the LMs found in four out of many energy functions used in this work: RBMs trained with 1, 5, 18 and 1400 epochs. In each case, the LMs here were found after the 5th (the last) cycle of many SA runs. The figure demonstrates that the different energy functions provide significant variety of the LM energies for investigating which LVs can be found by the D-Wave (and also, can be found only by the D-Wave), when QA and SA are applied to those energy functions.

Fig. 3 shows overlapping histograms of the barrier heights (the escape barriers $E_{act}$) of the LVs for three (out of four shown in Fig.2) energy functions: RBMs trained with 1, 5 and 1400 epochs. The figure demonstrates that the energy functions produced by training the RBM with a small number of iterations have a small number of very deep and a large number of shallow LVs. The energy functions corresponding to a more prolonged training become rather different; they have somewhat more homogeneous distribution of LVs, with a higher number of medium-size barriers.

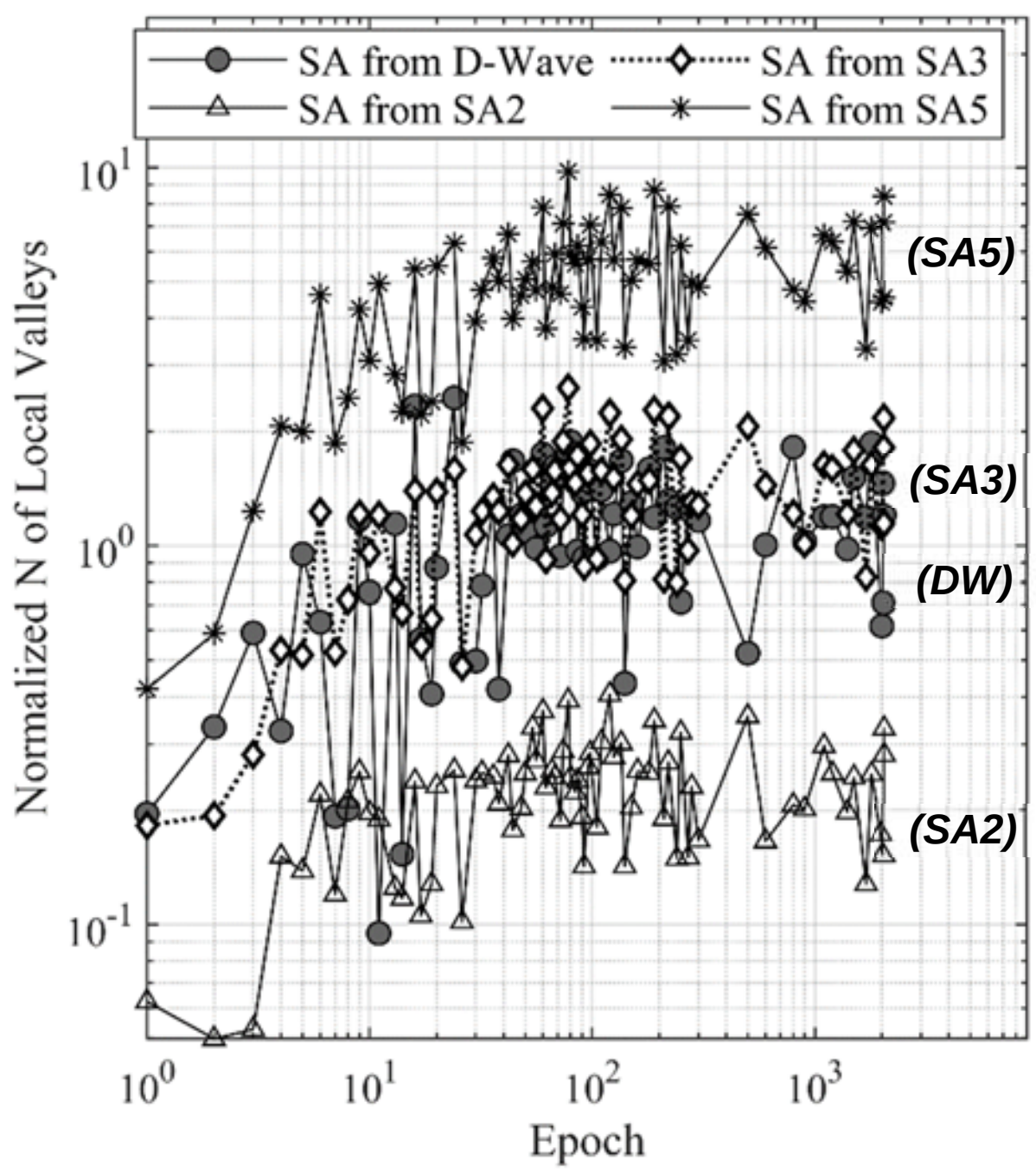

Fig.4 The number of LVs found by the D-Wave (DW), as a function of the training epoch, compared to that found by the classical simulated annealing (SA) with increasingly more MC cycles: (SA2) ~1x10$^5$ cycles, (SA3) ~1x10$^6$ cycles, and (SA5) ~1x10$^7$ cycles. The number of LV is normalized to the number of the patterns in the training dataset.

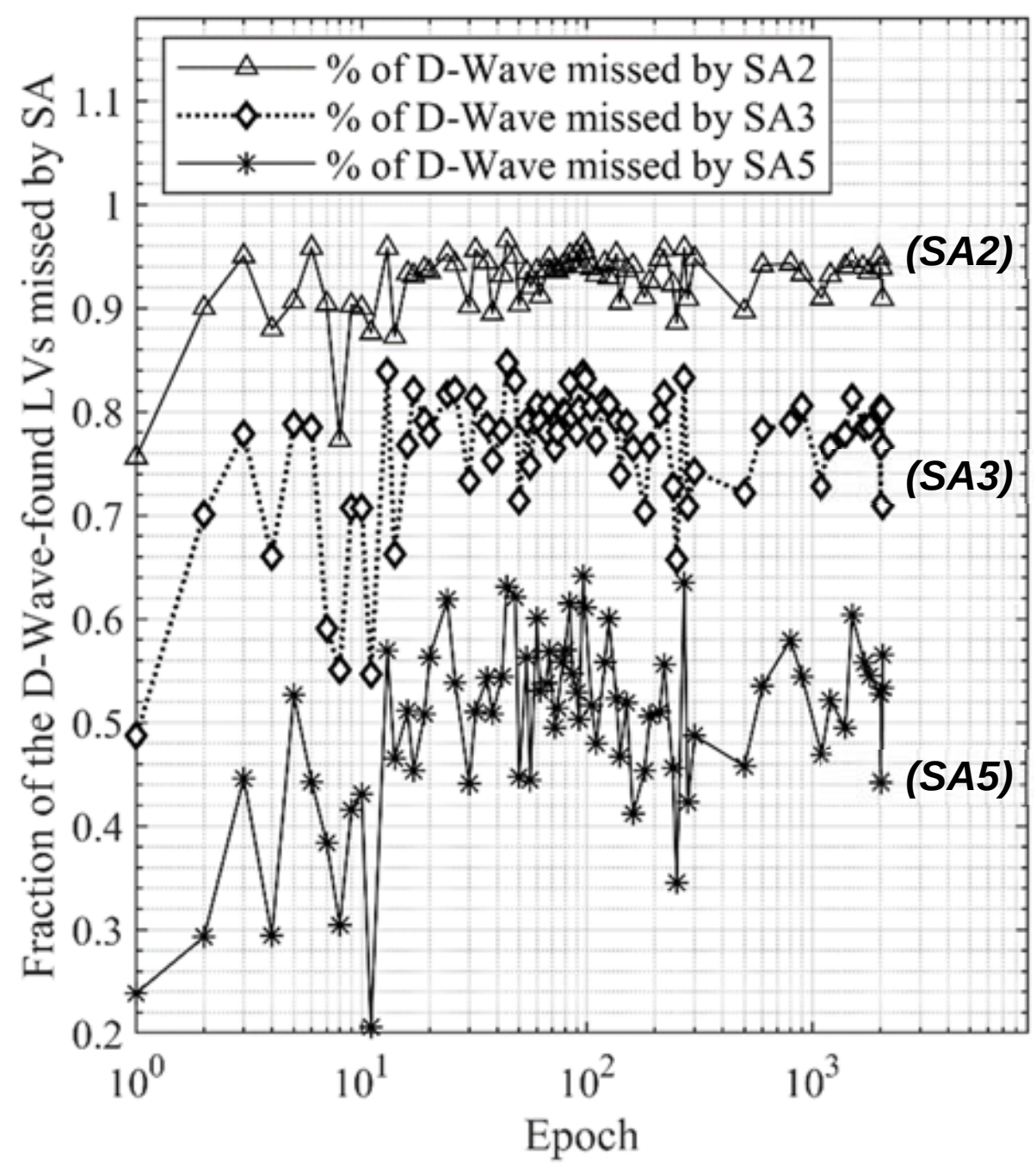

Fig.5 Percentage of the LVs found by the D-Wave that were missed by the classical SA, plotted as a function of the training epoch. The 3 curves shown are for the 3 cases of increasingly more MC cycles (as in Fig.4): (SA2) ~1x10$^5$ cycles, (SA3) ~1x10$^6$ cycles, and (SA5) ~1x10$^7$ cycles..

It was also established that the energy functions of the RBM after different training iterations possess a different number of LVs (or at least those LVs that get found). This number increases almost an order of magnitude when going from a single training epoch to approximately 100 training epochs and beyond (Fig.4).

It follows that different energy functions used in this work give a good variety for investigating which LVs (low or high energy, shallow or deep, etc.) get more easily found by the D-Wave (and also, found only by the D-Wave), when it is applied to energy functions having a very different number (from a few to many) of very diverse LVs.

*B. The number and the uniqueness of the LVs found by the two techniques.*

Next, for all the available training epochs and corresponding very different energy functions, we investigate how many LVs can be found by a single D-Wave call, and by SA with increasing number of annealing cycles up to a few weeks of continuous SA runs. Also, the main interest of this work was on what unique results not easily available from the classical searches come from the D-Wave. Therefore, we look specifically at the D-Wave found LVs, and compare the numbers of those that are found only by the D-Wave (i.e., missed by SA) and those found by both techniques.

In Fig.4, the number of LVs found by the D-Wave (labeled as (DW)) is shown as a function of the number of the training epochs and is compared to the number of LVs found by the classical simulated annealing (SA). The three curves are for three cases of increasingly higher number of MC cycles: (SA2) had ~$1x10^5$ cycles, (SA3) had ~$1x10^6$ cycles, and (SA5) had ~$1x10^7$ cycles. The number of LV is normalized to the number of the patterns in the training dataset. The local minima in Fig.4 curve (DW) have been found by relaxation at $T$=0 from each solution found by the single D-Wave call. The local minima in (SA2,3 and 5) have been found by an increasingly larger number of classical SA runs.

More than a week of continued SA runs (SA3) was required for SA to find approximately as many LV as found by the D-Wave. The longest SA (SA5) attempted in this work took multiple weeks of computer time, which allowed it find 4-5 times more LVs than the D-Wave.

The percentage of the LVs found by the D-Wave that were missed by the classical SA is plotted in Fig.5 as a function of the training epoch. Again, the three curves shown are for the three cases of increasingly more MC cycles (as in Fig.4). For SA3, when SA finds nearly as many LV as D-Wave (see Fig.4), >50-80% (depending on the epoch) QA-found LVs were not found by the classical SA.

For SA5, when SA finds 4-5 times more LV than D-Wave (see Fig.4), the expectation had been that the classical search should be able to find all the LVs found by the D-Wave. Instead, more than 30-50% of the QA-found LVs remained not found by SA even for this longest classical search, at least for the relatively small RBM graph that can be embedded in the current version of the QA hardware.

*C. Comparison of the LVs found by the D-Wave and SA*

The next question was about a relative importance of those LVs that are found by the D-Wave but missed by the classical search. A pessimistic expectation (with respect to the unique abilities of QAs) could be that the LVs found only by the D-

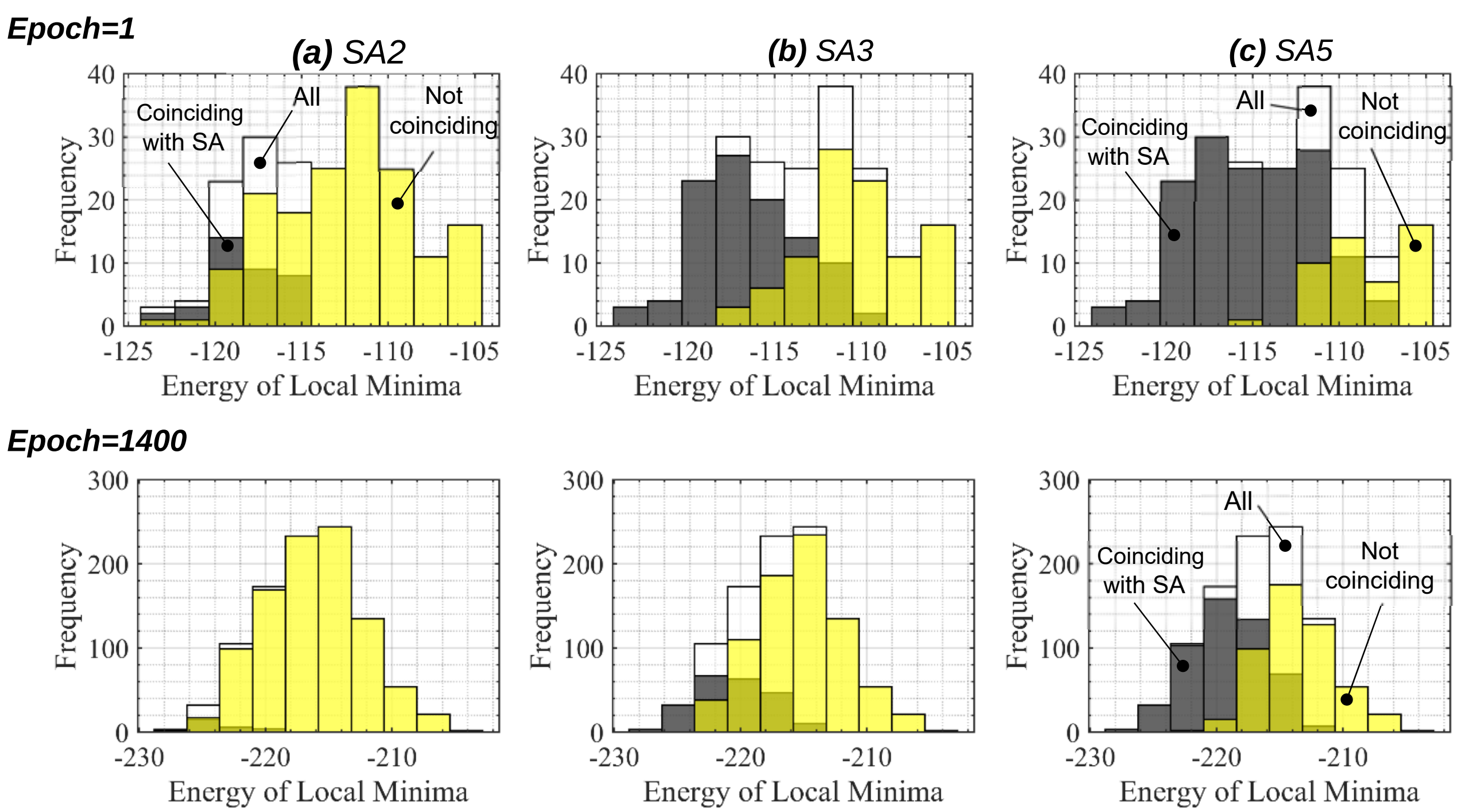

Fig.6 Histograms of the RBM energies of the LMs present in the D-Wave sample, plotted for two different energy functions corresponding to two training durations: (top) one epoch and (bottom) 1400 epochs. The white bars in each figure are for all LMs in the D-Wave sample. The dark (gray) bars are for those LMs in the D-Wave sample that coincide with the LMs found by the classical SA. The light (yellow) bars are for those D-Wave-found LMs that were not found by SA. Lighter shade of gray occurs where portions of the dark (gray) and light (yellow) bars happen to be covered by each other. The three columns correspond to increasingly more cycles of SA used in the classical part of the search: column (a) SA2 with ~$1x10^5$ cycles, column (b) SA3 with ~$1x10^6$ cycles and column (c) SA5 with ~$1x10^7$ cycles.

Wave would all turn out to have high energy (i.e., low probability of the corresponding states), low thermal activation energy, low width of the main (the lower) portion of the LV, which is where the system spends most of the time at temperatures having practical importance for different applications. The reality turned out to be much more promising.

Fig.6 shows histograms of the RBM energies of the LMs present in the D-Wave sample (found by relaxation at T=0 from each of the 1000 repetitions of the D-Wave solutions), plotted for two energy functions corresponding to two training durations: (top) one epoch and (bottom) 1400 epochs, which are the energy functions very different with respect to the number of LVs available and their properties. The dark (gray) bars are for those LMs in the D-Wave sample that coincide with the LMs found by the classical SA. The light (yellow) bars are for those D-Wave-found LMs that were not found by SA. Please note that a lighter shade of gray in Fig. 6 (as well as in Fig. 7) occurs where portions of the dark (gray) and light (yellow) bars happen to be covered by each other. The three columns correspond to increasingly more cycles of SA used in the classical search (as in Figs.4 and 5): column (a) SA1 with ~1x$10^5$ cycles, column (b) SA3 with ~1x$10^6$ cycles and column (c) SA5 with ~1x$10^7$ cycles. The white bars in each figure are for all LMs in the D-Wave sample.

Analyzing the results in Fig.6 from (a) to (c) for both energy functions (top and bottom), it follows that, for both of those very different energy functions, more persistent classical searches result in SA eventually finding all the LM having the lowest energies (dark/gray bars). The QA-found LM that remain not found by SA (light bars) even after multiple weeks of the classical search are predominantly at higher, but also at intermediate, energy.

From the point of view of this criterion alone (the energy of LMs), the LVs found only by the D-Wave could be expected to be less important compared to those found by both techniques, at least when the classical technique takes very large computational resources and time. However, when the classical search is reasonably long but not prohibitively long for many applications (days rather than weeks, Fig.6(a)), LVs found only by the D-Wave include also those at the lowest part of the energy histogram (i.e., the more stable states in the probability distribution).

Further, for many applications, it is not as much the energy of the LM as the height of the escape barrier that is important (e.g., for the comparative stability of the corresponding states). The barrier height as well as some other LV properties are investigated next.

Histograms of various properties of LVs present in the D-Wave sample are shown Fig.7. The LV properties are as defined in Section III.A, from the top to bottom: the barrier height ($E_{act}$), the *DOS* near the bottom of the LV, the width-related parameter of the LV (*W*), and the number of the sampled states inside LV ($N_{LV}$, comprising both $N_{low}$ and $N_{up}$). The three columns in Fig.7 represent three different energy functions: RBMs trained with 1, 5 and 1400 epochs. Again, each figure includes those QA-found LVs that coincide with the LVs found by the classical SA (dark/gray bars) as well as those that were not found SA (light/yellow bars). The results of the classical LV search in each figure are for SA3, which is when SA found nearly as many LV as D-Wave (see Fig.4). Jumping forward, it should be noted that (not shown) similar histograms plotted for SA5 (the longest classical search used in this work) communicated the same trend and conclusions.

For all the three energy functions, the histograms of $E_{act}$ (the first row in Fig.7) do not exactly support the expectation of high importance of the LVs that are found only by the D-Wave. All the LVs having the highest $E_{act}$ (the most stable states in the configuration space) are found by both techniques (the dark bars). The LVs found only by the D-Wave have medium to low values of $E_{act}$ (the light/yellow bars). A positive observation with respect to the $E_{act}$ criterion is that the number of those states in the high $E_{act}$ region without any light/yellow bars is relatively small – less than 20 in the examples of Fig.7. In contrast, there are many more light/yellow bars with lower $E_{act}$ that may be still high enough to make those LVs potentially important for many applications.

For the next two important criteria of potential LV importance – the *DOS* at the bottom of the LV and the width-related parameter *W* of the main (lower) portion of the LV (the second and the third rows of Fig.7) – the results are more favorable for the D-Wave. LVs found only by the D-Wave (the light/yellow bars) are distributed across the entire range of the values of *DOS* and *W*, which means that those LVs include the potentially important kind, having large *DOS* and *W*. Actually, it may be observed that the widest LVs and those with the largest *DOS* are dominated by those that were found only by the D-Wave, with smaller number of those found by both techniques. In other words, it is the widest LVs that are for some reason get missed by the classical search but found by the D-Wave. However, we do not have enough reasons to claim a universal nature of this (relatively moderate) trend or speculate about its mechanism.

Finally, we look at the "size" of the entire LV (the last row of Fig.7), which is expressed as the number of states $N_{LV}$ that can be sampled up to the the most remote states of the LV having energies close to $E_{max}$ from the bottom of the LV (see Fig.1).

It follows from the last row of Fig.7 that it is mostly smaller-to-medium LVs that are found by only the D-Wave, while the largest LVs are only those that were found by both techniques. The size of the LV here refers to the $N_{LV}$ (the number of states in the entire LV), not just the most important states $N_{low}$ in the most important lower portion of the LV.

It should be noted that most of the higher energy part of the LV's BoA represent states that probably are never visited by the system when it operates close to the equilibrium for most of the applications. Further, when considering reactions involving thermal jumps between LVs, it is $\Delta E_{act}$ rather than $E_{max}$ and $N_{low}$ rather than $N_{LV}$ or $N_{up}$ that are relevant. However, the BoA and its relative size must be important for the local search by SA to converge to the bottom of the corresponding LV. This possibility is verified next.

### *D. Possible reasons for the D-Wave and the SA finding different LVs often having similar properties.*

As reported in the previous section, many wide LVs and, actually, the widest LVs for some reason get missed by the classical search but found by the D-Wave. The trend for the $N_{LV}$ (the size estimate of the entire LV) in Fig.7 hints at a straightforward explanation of why certain LVs found by the D-Wave are missed by the classical search, even if those LVs have other parameters similar to some LVs found by both techniques. Large $N_{LV}$ could primarily reflect large BoA of the LVs, regardless of the size of the main lower portion of the LV ($N_{low}$ and $W$).

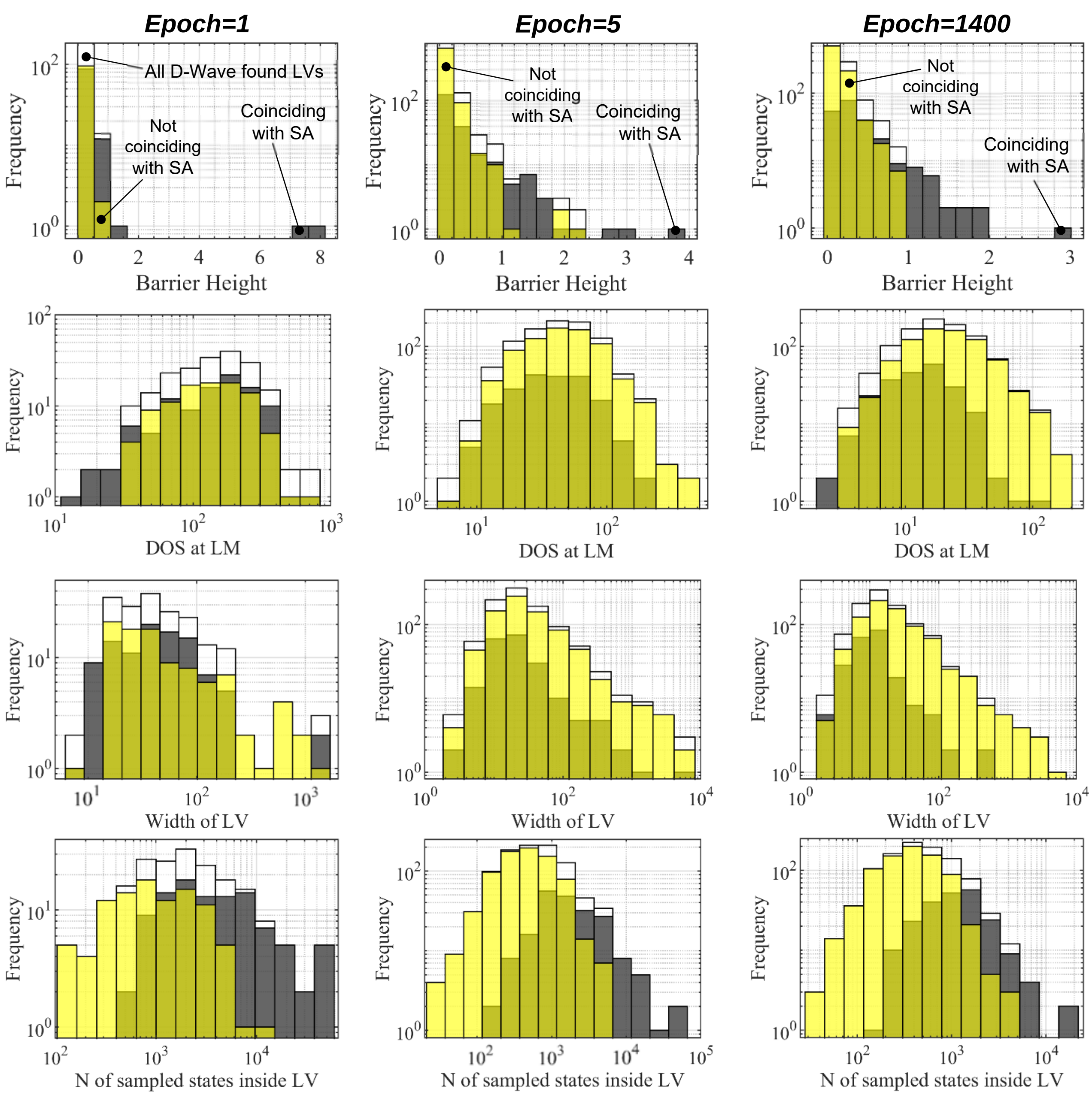

Fig.7 Histograms of various properties (defined in Section III.A) of LVs present in the D-Wave sample, from top to bottom: barrier height ($E_{act}$), *DOS* near the bottom of the LV, width of the LV ($W$), and number of the sampled states inside LV ($N_{LV}$). The three columns represent three different energy functions: RBMs trained with 1, 5 and 1400 epochs. Each figure includes all D-Wave-found LVs (white bars), those D-Wave-found LVs that coincide with the LVs found by the classical SA (dark/gray bars) as well as those that were not found by SA (the light/yellow bars). The results of the classical LV search in each figure are for SA3, which is when SA found nearly as many LV as D-Wave (see Fig.4).

To verify this, we look at the ratio of the upper states $N_{up.Coinc}$ and $N_{up.Not\ Coinc}$. Here, $N_{up.Coinc}$ is the number of the upper states (with energy above $E_{act}$) in those QA-found LV that coincide with SA-found LVs. $N_{up.Not\ Coinc}$ is the number of the higher-energy states in those QA-found LV that do not coincide with SA-found LV.

For four energy functions corresponding to 1, 5, 18 and 1400 training epochs, this ratio was found to be 4.6, 6.2, 3.5 and 2.7 respectively. This means that for all the energy functions investigated in this experiment, the size of the BoA is a few times larger in those QA-found LVs that could be also found by the classical search than in those LVs that were found only by the D-Wave. This can be considered as an expected result due to the fact that, when using a Monte Carlo based SA, with the MC chain initiated from random initial states in the configuration space, the finding of a particular LV is facilitated by the existence of a large number of BoA states providing a negative slope of the energy function towards the bottom of the given LV.

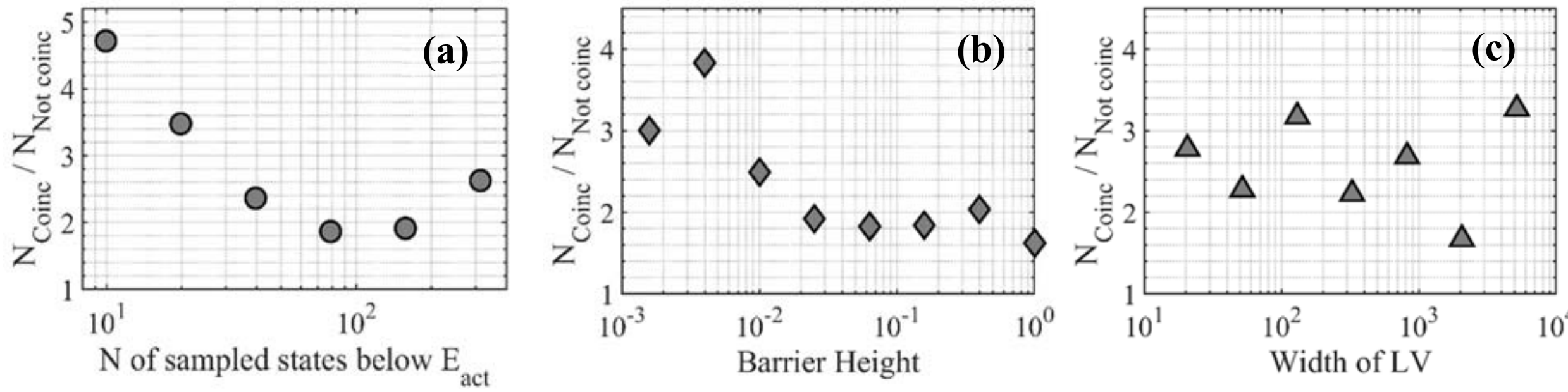

Fig.8 (a) The ratio of the upper states $N_{up.Coinc}$ and $N_{up.Not\ coinc}$, plotted for different values of the "size" of the LV below $E_{act}$. Here, $N_{up.Coinc}$ is the number of upper states (above $E_{act}$) in those D-Wave-found LV that coincide with SA-found LV. $N_{up.Not\ coinc}$ is the number of upper states in those D-Wave-found LV that do not coincide with SA-found LV. The "size" of the LV is estimated as $N_{low}$ - the number of states below $E_{act}$ sampled from each LV. (b) The same ratio plotted for different values of the barrier height (i.e., $E_{act}$). (c) The same ratio plotted for different values of the width-related parameter of the main low portion of the LV below $E_{act}$. The results are for epoch = 1400 and the longest classical SA search, SA5.

The final question was if this $N_{up.Coinc}$ / $N_{up.Not\ coinc}$ criterion is equally relevant in a wide range of important LV properties. Fig.8 (a) shows $N_{up.Coinc}$ and $N_{up.Not\ coinc}$, plotted for different values of the "size" of the LV below $E_{act}$. Again, the "size" of the LV was estimated as $N_{low}$ - the number of states below $E_{act}$ sampled from each LV. The results in Fig.8 are for the 1400 training epochs and for the longest classical SA search SA5.

Fig.8(b) and (c) show the same ratio $N_{up.Coinc}$ / $N_{up.Not\ Coinc}$ plotted respectively for different values of the barrier height $E_{act}$ and for different values of the width-related parameter W of the main low portion of the LV.

The figures demonstrate that for large or small, shallow or deep, wide or narrow LVs (when considering the most important part of the LV close to its bottom and below $E_{act}$), the LV found by the D-Wave but missed by the most prolonged classical SA search in this work are distinguished by a few-times smaller size of the LV's BoA characterized by $N_{up}$, the number of upper states above $E_{act}$. In extreme cases, it means that even relatively wide LVs with a large number of low-energy states and high escape barriers (i.e., rather stable states) may be missed by the classical search when those LVs do not have large enough BoA - the upper states with a negative gradient in the direction of the corresponding LV bottom. An important result is that, as confirmed by the $N_{up.Coinc}$ / $N_{up.Not\ Coinc}$ trend in Fig.8, that property of LVs (the size of the BoA) is not relevant or is at least less relevant for the QA search, allowing QA to easily find some potentially important (wide and deep) LVs missed by even prohibitively lengthy classical searches.

## V. Conclusion

As was reported earlier [12][13], a single call to the D-Wave normally misses many LVs that can be found by prolonged SA. However, since we already know how to find those LVs classically, an important focus of this work was on those LVs that could not be found classically when using a reasonable amount of computational resources and time but are found by the D-Wave relatively easily.

At least for the energy functions investigated in this work, which have a large number of "findable" LVs, it takes the classical search unproportionably more time to find the same number of LVs as that found by the D-Wave in a fraction of a second (1000 results within tens of milliseconds). Naturally, an infinitely long classical search could eventually find all the existing LVs. In practice, the efforts invested in the prohibitively long (for most applications) search attempted in this work allowed us to find 4-5 times more LVs than a single D-Wave call. And yet, even this most extensive classical search failed to find 30-50% of the QA-found LVs, depending on the energy function investigated.

A proof was obtained for the fact that at least some of the LVs found by the D-Wave but missed by the classical search satisfy a number of criteria that are likely to make them important for scientific and engineering applications. Those criteria used when comparing LVs found by the D-Wave only and found by both techniques were selected (1) from the point of view of potential relevance for typical applications in ML, physics, chemistry, materials science and engineering that benefit from finding difficult to find LVs and (2) from the point of view of providing computational feasibility of extracting relevant properties of many LVs in a realistic time. Also, comparison of those LV properties for different number of graphical model training iterations provided reassurance that the conclusions of this work should apply also to energy functions that are very different with respect to the number of LVs, their energies, etc.

Somewhat disappointingly, the D-Wave showed no "hero results" with respect to finding LVs missed by the classical search that are specifically at the highest end of the energy of the LM and the activation energy for escaping form the LV. The likely ground state (GS), a few states with energies close to the GS, as well as (or simultaneously) the most thermally stable states are normally found by both techniques if the classical search is long enough. However, excluding this relatively small number of the lowest-energy and/or the deepest LVs (compared to the total number of the LVs found by the D-Wave), the D-Wave finds a reasonable number of LVs in the intermediate range of those parameters, which may be important for many applications, such as, for example, search for unknown metastable states.

The D-Wave showed positive results with respect to other LVs parameters used in the comparison. When considering the most important lower part of the LV, which contains states in which systems would spend most of the time in most practical applications, the LVs found only by the D-Wave included some having the highest width and the highest *DOS* at the bottom of the LV.

Finally, an evidence was obtained in support of a straightforward explanation for the reason of why some relatively wide LVs found by the D-Wave are missed by the classical search. A bigger size of the BoA of the LVs seems to be important for ensuring that the LV is not easily missed by the classical search. In contrast, the size of the BoA seems to be much less important for the QA search, allowing the D-Wave to easily find some potentially important (wide and deep) LVs that are likely to be missed by classical searches.

An open question is whether this advantage of the QA persists across graphs other than RBMs, graphs bigger than the ones used in this work as well as for many different applications (that require temperatures low enough for the concept of the found versus missed LVs to be important).

The immediate future work will concentrate on the LVs found only by the D-Wave and having most favorable values of the LV parameters investigated in this work. The goal will be to establish what those states are and how important they could be for specific sorts of applications beyond sampling in ML.

## Acknowledgment

The authors thank D-Wave Systems for access to their 2000 Q machine.

## References

[1] D-Wave Systems, Inc, http://www.dwavesys.com

[2] Mott A, Job J, Vlimant J R, Lidar D and Spiropulu M 2017 Nature 550 375{379 URL https://doi.org/10.1038/nature24047

[3] Ushijima-Mwesigwa H, Negre C F A and Mniszewski S M, 2017 Proceedings of the Second International Workshop on Post Moores Era Supercomputing PMES'17 (ACM) pp 22{29 ISBN 978-1-4503-5126-3 URL http://doi.acm.org/10.1145/3149526.3149531

[4] Li R Y, Felice R D, Rohs R and Lidar D A 2018, NPJ Quantum Information 4, URL https://doi.org/10.1038/s41534-018-0060-8

[5] O'Malley D, Vesselinov V V, Alexandrov B S and Alexandrov L B 2018, PLOS ONE 13 1-12 URL https://doi.org/10.1371/journal.pone.0206653

[6] Rose, G. (2014). First ever DBM trained using a quantum computer. https://dwave.wordpress.com/2014/01/06/first-ever-dbm-trained-using-a-quantum-computer/

[7] Adachi, S.H., & Henderson, M.P. (2015). Application of Quantum Annealing to Training of Deep Neural Networks. eprint: arXiv:1510.06356.

[8] Benedetti, M., Reaple-Gómez, J., Biswas, R., & Perdomo-Ortiz, A. (2015). Estimation of effective temperatures in a quantum annealer and its impact in sampling applications: A case study towards deep learning applications, Physical Review A 94, 022308 (2016)

[9] M. Benedetti, J. Realpe-Gómez, R. Biswas, and A. Perdomo-Ortiz, "Quantum-Assisted Learning of Hardware-Embedded Probabilistic Graphical Models," Phys. Rev. X 7, 041052

[10] M. H. Amin, E. Andriyash, J. Rolfe, B. Kulchytskyy, and R. Melko, Quantum Boltzmann Machine, arXiv:1601.02036

[11] G. Carleo and M. Troyer, "Solving the quantum many-body problem with artificial neural networks,"Science, vol. 355, p. 602-606 (2017), published Feb 10, 2017

[12] Y. Koshka and M. A. Novotny, "Towards Sampling from Nondirected Probabilistic Graphical models using a D-Wave Quantum Annealer," arXiv:1905.00159

[13] Y. Koshka and M. A. Novotny, "Comparison of Use of a 2000 Qubit D-Wave Quantum Annealer and MCMC for Sampling, Image Reconstruction, and Classification," IEEE Transactions on Emerging Topics in Computational Intelligence., doi: 10.1109 /TETCI.2018.2871466

[14] Y. Koshka, D. Perera, S. Hall and M. A. Novotny, "Determination of the Lowest-Energy States for the Model Distribution of Trained Restricted Boltzmann Machines using a 1000 Qubit D-Wave 2X Quantum Computer," Neural Computation 29, 1815–1837 (2017).

[15] Dueck, G. and Scheuer, T. (1990) Threshold accepting—a general-purpose optimization algorithm appearing superior to simulated annealing. *Journal of Computational Physics,* **90**, 161–175.

[16] Charon, I. and Hudry, O. (1993) The noising method—a new method for combinatorial optimization. *Operations Research Letters,* **14**, 133–137.

[17] Glover, F. (1994) Tabu search for nonlinear and parametric optimization (with links to genetic algorithms). *Discrete Applied Mathematics,* **49,** 231–255.

[18] Liepins, G.E. and Hilliard, M.R. (1989) Genetic algorithms: foundations and applications. *Annals of Operations Research,* **21,** 31–58.

[19] Henderson D., Jacobson S.H., Johnson A.W. (2003) The Theory and Practice of Simulated Annealing. In: Glover F., Kochenberger G.A. (eds) Handbook of Metaheuristics. International Series in Operations Research & Management Science, vol 57. Springer, Boston, MA

[20] Santoro, G. E., & Tosatti, E. (2006). Topical Review: Optimization using quantum mechanics: quantum annealing through adiabatic evolution. J. Phys. A: Math. Gen., **39.** pp. R393–R431.

[21] K. Binder and A. P. Young, Spin glasses: Experimental facts, theoretical concepts and open questions, Rev. Mod. Phys. 58, (1986), p. 801

[22] Stein, D. L., & Newman, C.M. (2013). Spin Glasses and Complexity. Princeton University Press, Princeton, NJ.

[23] Boixo, S., T. F. Rønnow, T. F., Isakov, S. V., Wang, Z., Wecker, D., Lidar, D. A., Martinis, J. M., & Troyer, M. (2014). Evidence for quantum annealing with more than one hundred qubits. Nature Physics, Volume 10, Issue 3, pp. 218-224.

[24] Trummer, I., & Koch, C. (2015). Multiple Query Optimization on the D-Wave 2X Adiabatic Quantum Computer. arXiv:1510.06437

[25] Novotny, M.A., Hobl, L., Hall, J.S., & Michielsen, J.S. (2016). Spanning Tree Calculations on D-Wave 2 Machines. Journal of Physics: Conference Series, vol. 681, 012005. International Conference on Computer Simulation in Physics and Beyond (CSP 2015), 6-10 Sept. 2015, Moscow, Russia, IOP Publishing Ltd., 2016. https://iopscience.iop.org/article/10.1088/1742-6596/681/1/012005/meta

[26] Fischer, A., & Igel, C. (2014). Training restricted Boltzmann machines: An introduction. Pattern Recognition, Volume 47, Issue 1, January 2014, pp. 25-39.

[27] M.A. Novotny, *Monte Carlo Algorithms with Absorbing Markov Chains: Fast Local Algorithms for Slow Dynamics,* Phys. Rev. Lett. **75**, 1-5 (1995); erratum Phys. Rev. Lett. **75**, 1424 (1995)

[28] M. Kolesik, M.A. Novotny, and P.A. Rikvold, *Projection Method for Statics and Dynamics of Lattice Spin Systems,* Phys. Rev. Lett. **80**, 3384 (1998).

[29] K. Biswas and M A Novotny, *Mapping the dynamics of multi-dimensional systems onto a nearest-neighbor coupled discrete set of states conserving the mean first-passage times: a projective dynamics approach,* Journal of Physics A: Mathematical and Theoretical, Volume 44, (2011)

[30] W. E. and E. Vanden-Eijnden. Towards a theory of transition paths. *J. Stat. Phys.*, vol. 123, No. 3, 503-523, 2006.

[31] W. E., W. Ren and E. Vanden-Eijnden. String method for the study of rare events. *Phys. Rev. B*, vol. 66, no. 5, 052301, 2002.

[32] W. E., W. Ren, E. Vanden-Eijnden. Simplified and improved string method for computing the minimum energy paths in barrier-crossing events. *J. Chem. Phys.*, vol. 126, no. 16, 164103, 2007.

**Yaroslav Koshka** received his B.S. and M.S. in Electronics in 1993 from Kiev Polytechnic University, Kiev, Ukraine, and his PhD in electrical engineering in 1998 from the University of South Florida. From 1993 till 1995, Dr. Koshka worked as an Engineer Mathematician at the Institute for Problems of Material Science, Kiev, Ukraine. From 1998 till 2002, he was a postdoctoral fellow and an Assistant Research Professor at Mississippi State University (MSU), working predominantly in experimental micro- and nano-electronics. He joined the faculty at MSU in 2002. He is currently a Professor in the Department of Electrical and Computer Engineering at MSU and the director of the Emerging Materials Research Laboratory. His current main research area is quantum computations, their application to machine learning as well as to properties of electronic materials. Other research interests include semiconductor materials and device characterization, defect engineering, synthesis of wide-bandgap semiconductor materials and nanostructures, physics of semiconductor devices, and nanoelectronics.

**Mark A. Novotny** received his B.S. in 1973 from North Dakota State University, Fargo, North Dakota and Ph.D. in 1979 from Leland Stanford Junior University, Stanford, California, both in physics.

Since 2001 he has been Professor and Head of the Department of Physics and Astronomy at Mississippi State University, Mississippi State, Mississippi. He has published more than 200 papers in refereed journals. His research area is classical and quantum computational physics, with applications to classical and quantum properties of materials and statistical mechanics.

Prof. Novotny has the honor of being a Giles Distinguished Professor at Mississippi State University, as well as being a Fellow of the American Physical Society and a Fellow of AAAS.